\documentclass[onecolumn, times, trackchanges]{aastex63}

\usepackage{txfonts}
\newcommand{\target}{KIC 12602250 }

\begin{document}

\title{\target: A low-amplitude Double-mode $\delta$ Scuti star with Amplitude Modulation}
\shorttitle{$Kepler$ Observations of \target}
\shortauthors{Lv, C. et al.}

\author{Chenglong Lv}
\affil{Xinjiang Astronomical Observatory, Chinese Academy of Sciences, Urumqi, Xinjiang 830011, People's Republic of China}
\affil{School of Astronomy and Space Science, University of Chinese Academy of Sciences, Beijing 100049, People's Republic of China}

\author{Ali Esamdin}
\email{aliyi@xao.ac.cn}
\affil{Xinjiang Astronomical Observatory, Chinese Academy of Sciences, Urumqi, Xinjiang 830011, People's Republic of China}

\author{Xiangyun Zeng}
\email{zengxiangyun@xao.ac.cn}
\affil{Xinjiang Astronomical Observatory, Chinese Academy of Sciences, Urumqi, Xinjiang 830011, People's Republic of China}
\affil{School of Astronomy and Space Science, University of Chinese Academy of Sciences, Beijing 100049, People's Republic of China}

\author{J.Pascual-Granado}
\affil{Instituto de Astrof\'isica de Andaluc\'ia - CSIC, 18008 Granada, Spain}

\author{Taozhi Yang}
\affil{School of Physics, Xi'an Jiaotong University, Xi'an 710049, People's Republic of China}

\author{Junhui Liu}
\affil{Department of Astronomy, Xiamen University, Xiamen, Fujian 361005, China}

\begin{abstract}

We propose for the first time that \target is a low-amplitude radial double-mode $\delta$ Scuti star with amplitude modulation. The detailed frequency analysis is given for the light curve of \target which is delivered from the Kepler mission. The Fourier analysis of the long cadence data (i.e. Q0 - Q17, spanning 1471 days) reveals that the variations of the light curve are dominated by the strongest mode with frequency F0 = 11.6141 $\rm{d^{-1}}$, suggesting that \target is a $\delta$ Scuti star. The other independent mode F1 = 14.9741 $\rm{d^{-1}}$ is newly detected. The amplitude of the light variations of \target is $\sim$ 0.06 mag, which indicates that this is a low-amplitude $\delta$ Scuti star, but the ratio of F0/F1 is estimated as 0.7756 which is typical of HADS, and a slow amplitude growth is detected in F1 and $f_{3}$, which could be due to stellar evolution, suggesting that \target could be a post main sequence $\delta$ Scuti which is crossing the instability strip for the first time.

\end{abstract}


\section{Introduction}

The high-precision photometric data provided by Kepler provides an unprecedented opportunity to explore stellar interiors by using the natural oscillation mode of stars, thus greatly expanding the research field of asteroseismology \citep{{Chaplin2010}}. The ultra-high precision photometric observations at $\mu$ mag level have greatly improved our understanding of many types of pulsating variables \citep{2012MNRAS.419.3028B}. \citet{Bedding2011} proposed that the observed period spacings of gravity modes could be applied to distinguish the hydrogen and helium burning stars in red giants, while \citet{2018Natur.554...73G} suggested that there may be a suppressed oxygen-dominated core in pulsating white dwarfs. As a group of conventional variable stars, the intrinsic homogenous pulsating modes of $\delta$ Scuti stars make them excellent targets in the study of asteroseismology \citep{Yang2018a,2011MNRAS.414.1721B,2011MNRAS.417..591B}. The fundamental mode, first, second, even the third and fourth radial pulsation mode could be indicators to the interiors burning mechanisms of $\delta$ Scuti stars \citep{Breger2000}.

The $\delta$ Scuti type pulsating stars lie inside the classical Cepheid instability strip main sequence on the Hertzsprung Russell (HR) diagram. $\delta$ Scuti stars typically range from A2 to F2 in spectral type with luminosity classes from III to V (e.g., \citealp{Breger2000,1990A&AS...83...51L,2001A&A...366..178R}), and within the effective temperature range of 6300K $\leq$ T$_{\rm eff}$ $\leq$ 8600 K \citep{2011A&A...534A.125U}. The pulsating amplitudes are in the range of 0.003-0.9 mag in the V band, with periods usually between 0.02 and 0.3 d \citep{1979PASP...91....5B}. Some stars show amplitude modulation of pulsation modes caused by different reasons, e.g., beating, mode coupling, rotation, and Blazhko effect (\citealp{2014MNRAS.444.1909B,2016MNRAS.460.1970B,Yang2018b,Yang2019}). These targets are excellent samples for asteroseismic study, as they could improve our knowledge of the stellar structure and evolution of stars.

As a subclass of $\delta$ Scuti stars, the HADS stars usually pulsate with a light amplitude larger than 0.3 mag and generally rotate slowly with $v$ sin $i$ $\leq$ 30 kms$^{-1}$ \citep{Breger2000}. Compared with the low-amplitude $\delta$ Scuti stars, the HADS stars possess a more restrictive instability strip with a width in temperature of about 300 K and tend to shift to a lower temperature with evolution \citep{2000ASPC..210..373M}. \citet{2008PASJ...60..551L} reveal that only $\sim$ 0.24 percent of the stars suited in the $\delta$ Scuti region belong to HADS stars. The majority of HADS stars are typically young and metal-rich Population I stars; some have been confirmed to be SX Phe variables, and are Population II metal-deficient stars \citep{Breger2000,2012MNRAS.426.2413B}. In general, the HADS stars pulsate with only one or two modes (e.g., AE UMa, \citealp{2017MNRAS.467.3122N}; YZ Boo, \citealp{Yang2018a}; etc.), and most of their pulsations belong to radial modes.

\target is classified as a $\delta$ Scuti star with a pulsation period of 2.07 hrs by \cite{{2011A&A...529A..89D}}. According to the frequency analysis, we propose for the first time that \target is a low-amplitude radial double-mode $\delta$ Scuti star with no non-radial modes. The frequency spectra of \target is shown in Figure 2 while the basic parameters are tabulated in Table 1.

In this paper, the observations of \target are introduced in Section 2. Section 3 presents the frequency analysis while Section 4 discusses the radial triple modes and amplitude modulation of this star. We summarize in Section 5.
\begin{deluxetable}{lcc}
\tabletypesize{\small}
\tablewidth{0pc}
\tablenum{1}
\tablecaption{\target observational (photometry) data characteristics \label{Tab1}}
\tablehead{
\colhead{Parameters} &
\colhead{Value in \href{https://kasoc.phys.au.dk/catalog/12602250}{Catalog}}  &
\colhead{ }
}
\startdata
Kepler ID     & 12602250  \\
2MASS ID      & J19213193+5140320  \\
Gaia  ID      & 2139231542654343552  \\
RA,~Dec       & +19$^{h}$:21$^{m}$:31.9$^{s}$, +51$\degr$:40$\arcmin $:32.1$\arcsec$  \\
BJD$_\mathrm{0}$      & 2454953.5388                                   \\
Rayleigh$~f_\mathrm{res}$& 0.000679                                     & d$^{-1}$\\
Period                & 2.07                                      & hr\\
Kmag                  & 13.277                                    &   \\
Contamination         & 0.024                                      &     \\
$T_\mathrm{eff}$      & 6879                                           & K\\
$\log{g}$             & 4.131                                          & cgs\\
$\frac{R}{R_{\sun}}$  & 1.597                                           \\
$\frac{Fe}{H}$        & -0.138                                          & cgs\\
   \enddata
   \tablecomments{These parameters are available in the KASOC: https://kasoc.phys.au.dk/. }
\end{deluxetable}

\section{OBSERVATIONS AND DATA REDUCTION}

\begin{figure*}
\begin{center}
  \includegraphics[width=0.87\textwidth]{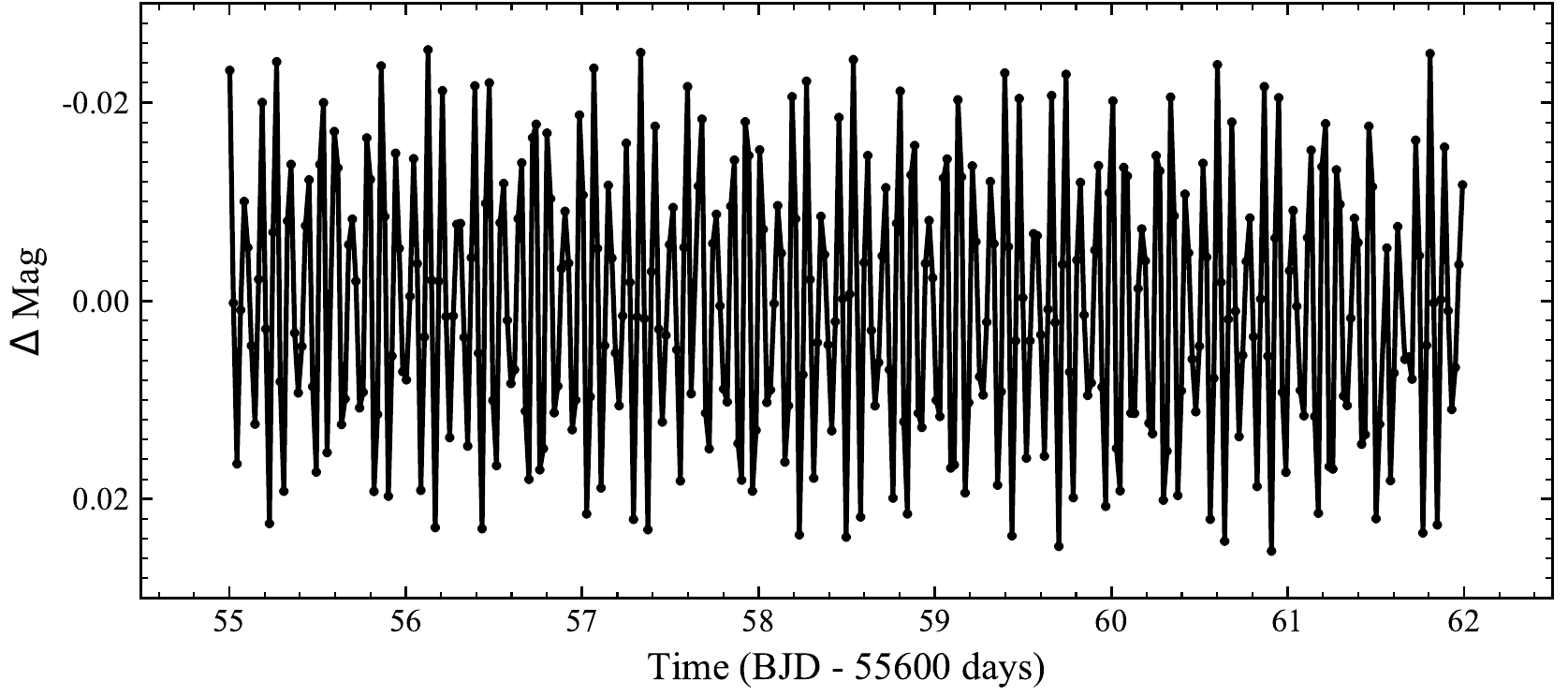}
  \caption{A portion of the long cadence light curve of \target. The amplitude of the light curve is about 0.06 mag.}
    \label{fig:light curve}
\end{center}
\end{figure*}

\target was observed from BJD 2454953.538 to 2456424.001 by $Kepler$ space telescope, including eighteen quarters (i.e. Q0-Q17). There are only LC photometric observations of \target available through the Kepler Asteroseismic Science Operations Center (KASOC) database$^{4}$ \citep{Kjeldsen2010} with two types: one is the raw flux, which is reduced by the NASA Kepler Science pipeline, and the other is corrected flux, which is provided by KASOC Working Group 4 (WG \# 4: $\delta$ Scuti targets). The second type has been corrected for systematic errors such as the cooling down, cooling up, outliers, and jumps. We use the corrected flux and convert it to magnitudes. Then the mean value of each quarter is subtracted to obtain the rectified time series. The final rectified light curve was obtained with 65264 data points, spanning over about 1471 days. Due to the gaps in time series cause spurious frequencies in the power spectra, in light curves of pulsating stars, this hampers identifying the theoretical oscillation modes. We used a forward-backward predictor based on autoregressive moving-average modeling (ARMA) in the time domain. The algorithm MIARMA is particularly suitable for replacing invalid data \citep{2015A&A...575A..78P,2018A&A...614A..40P}. We have used this algorithm to fill the gaps in the light curve of KIC 12602250. Figure \ref{fig:light curve} shows a portion of the rectified light curve of \target covering 7 days. From this figure, the peak-to-peak amplitude of \target obtained from the rectified light curve is $\sim$0.06 mag, suggesting that this star belongs to the low-amplitude class of $\delta$ Scuti stars.

\section{FREQUENCY ANALYSIS}

The software PERIOD04 \citep{Lenz2005} is employed to analyze the pulsating behavior of \target. The rectified light curve was fitted with the following formula:
\begin{equation}
m = m_{0} + \sum_{\mbox{\scriptsize\ $i$=1}}^N\mathnormal{A}_{i}sin(2\pi(\mathnormal{f}_{i}\mathnormal{t} + \phi_{i})), \label{equation1}
\end{equation}
where $m_{0}$, $A_{i}$, $f_{i}$, and $\phi_{i}$ are zero-point constant, amplitude, frequency and the corresponding phase, respectively.

Nyquist frequency sets the minimum sampling frequency that needs to be defined to prevent alias frequencies, while some potential significant frequencies could beyond Nyquist frequency \citep{Yang2018a,2011MNRAS.414.2413S,2016MNRAS.460.1970B}. The Nyquist frequency of LC observations is $f_{N}$ = 24.469 d$^{-1}$ \citep{2013MNRAS.430.2986M,2014MNRAS.439.2078H}. In order to detect all potential significant frequencies, we choose the frequency range is 0 $<$ $\nu$ $<$ 50 d$^{-1}$, which is somewhat wider than the typical pulsation frequency of $\delta$ Scuti stars \citep{1979PASP...91....5B,Breger2000,2011A&A...534A.125U,2011MNRAS.417..591B}. As LC observations apply longer integration time, the data signal-to-noise of LC observations is much higher than that of SC observations. So that the noise in the LC spectrum is lower than that in the SC observations. \citep{2010ApJ...713L.160G,Yang2018a}. Owing to the combined effects of a high signal-to-noise and a non-regular sampling in LC light curves, we might detect signal beyond the Nyquist limit and discriminate it from aliases.

\begin{figure*}
\begin{center}
  \includegraphics[width=0.87\textwidth]{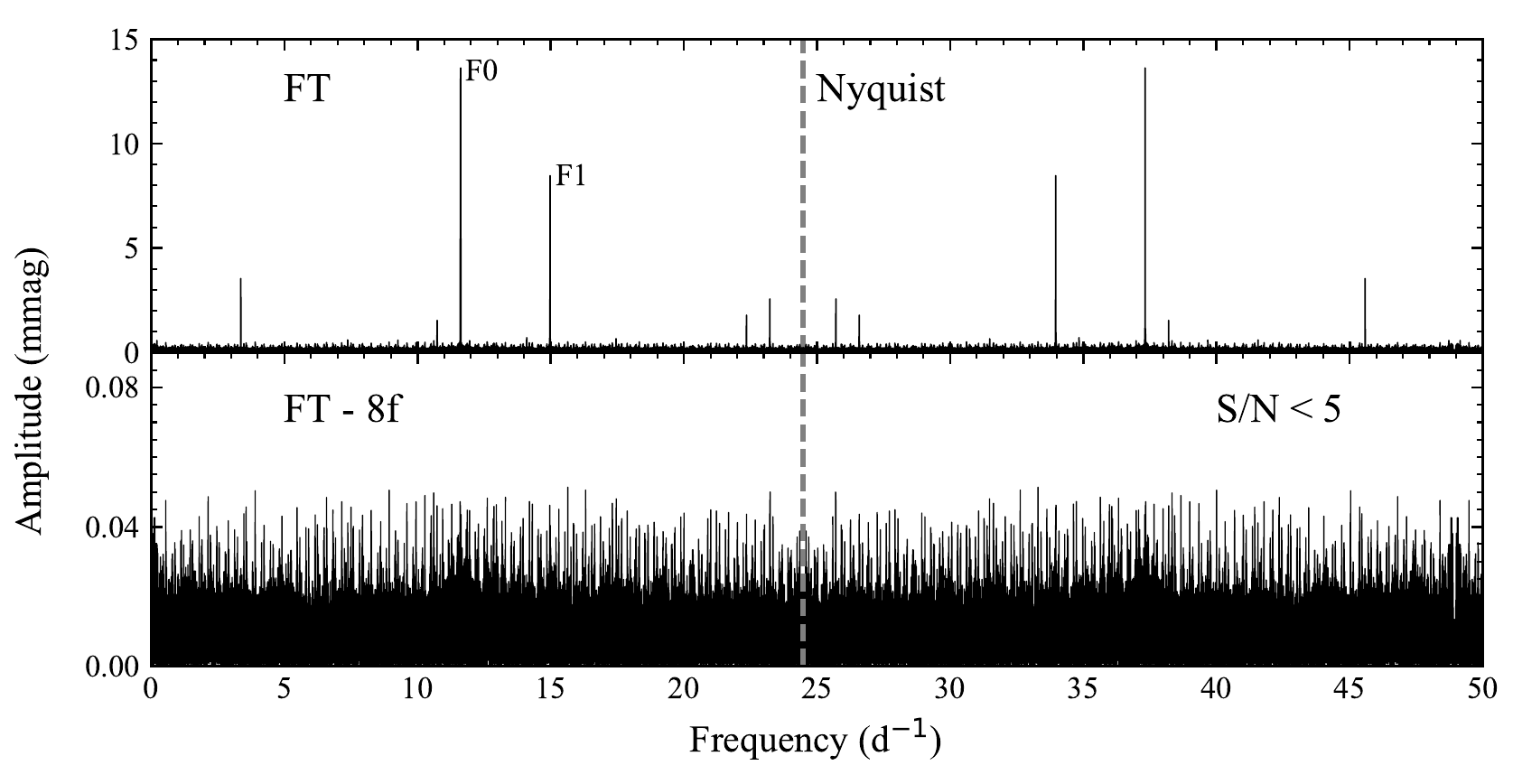}
  \caption{Fourier amplitude spectra and the prewhitening process for the light curve of \target. The top panel shows two independent frequencies, two harmonics and three combination frequencies. The independent frequencies F0, F1 are marked. The bottom panel shows the residual after subtracting 8 significant frequencies and the S/N $<$ 5. The dashed line refers to the Nyquist frequency.}
    \label{fig:amplitude spectra}
\end{center}
\end{figure*}

We consider two frequencies to be resolved if the difference between them is larger than the resolution frequency, $f_{res}$ = 1 / $T$, which is 0.000679 $d^{-1}$ for the LC light curve. The highest peak is usually identified as a potential significant frequency while extracting significant frequencies. Then a multi-frequency least-square fit using Equation 1 applies to the light curve with all significant frequencies detected and obtained the solutions for all the frequencies. A theoretical light curve constructed using the above solutions is subtracted from the rectified data while the residual is obtained for the next search. The above steps are repeated until no significant peak is detected in the frequency spectrum. The criterion of S/N $>$ 5.0 suggested by \cite{2015MNRAS.448L..16B} is adopted to determine the significance of the detected peaks. The uncertainties of frequencies were obtained following the method proposed by \citet{1999DSSN...13...28M}. Figure \ref{fig:amplitude spectra} shows the amplitude spectra and the pre-whitening procedures of the light curve. The last panel shows the residual amplitude spectra after pre-whiten eleven detected frequencies. No significant peak could be detected in the residual spectrum, which displays an overall distribution of typical noise.
\begin{deluxetable}{cccccc}
\tabletypesize{\small}
\tablewidth{0pc}
\tablenum{2}
\tablecaption{Extracted frequencies in LC data of \target \label{Tab2}}
\tablehead{
\colhead{$f_{i}$}   &
\colhead{Frequency (d$^{-1}$)}  &
\colhead{Amplitude (mmag)}      &
\colhead{S/N}            &
\colhead{Comment} &
}
\startdata
1&	11.6141(1)&	 13.695(6)     &	 498.0   & F0\\
2&	14.9741(2)&	 8.173(6)      &	 829.4   & F1\\
3&	3.3600(3) &	 4.874(6)      &	 62.6    & F1-F0\\
4&	23.2282(4)&	 2.697(6)      &	 38.3    & 2F0\\
5&	26.5883(6)&	 2.403(6)      &	 30.9    & F0+F1\\
6&	38.1998(6)&	 1.371(6)      &	 11.8    & F0+2F1\\
7&  7.3734 (4)&  0.612(6)      &     5.8     & 2$f_{N}$-F0-2F1\\
8&	29.9483(7)&  0.473(6)      &	 5.6     & 2F1\\
   \enddata
    \tablecomments{Among these frequencies, three peaks are independent frequencies, others are harmonic or combinations (denoted by $f_{i}$). $f_{7}$ is a alias frequency.}
\end{deluxetable}

A total of eight significant frequencies are detected in the spectra of \target and a full list is given in Table \ref{Tab2}. Among these frequencies, two of them are considered to be independent. It is reasonable that the strongest peak $f_{1}$ was assumed to be the fundamental mode, since the light variations were dominated by this frequency. Therefore, we marked $f_{1}$ with 'F0' in the last column of Table \ref{Tab2}, the other independent frequency $f_{2}$ is labeled as 'F1'. In addition, harmonics (i.e., $f_{4}$, $f_{8}$) of 'F0','F1' and three combination frequencies (i.e., $f_{3}$, $f_{5}$, $f_{6}$) of 'F0', 'F1' are also detected. The frequency $f_{7}$ consider as a alias frequency.

\section{DISCUSSION}

\subsection{The Double-mode in \target}

From a large number of stars, it was demonstrated that $\delta$ Scuti stars pulsate in low amplitude non-radial modes in addition to high-amplitude radial modes \citep{2016PhDT.......267B}. However, low amplitude $\delta$ Scuti stars, with only the radial pulsation mode are not common.
%

\citet{{Stellingwerf1979}} statistically analyzed the pulsation modes of a number of $\delta$ Scuti stars and presented the period ratios of the first four radial modes as:
$P_{1}$ / $P_{0}$ = (0.756 - 787), $P_{2}$ / $P_{0}$ = (0.611 - 632) and $P_{3}$ / $P_{0}$ = (0.500 - 525), in which $P_{0}$, $P_{1}$, $P_{2}$ and $P_{3}$ represent the fundamental mode, first overtone, second overtone and third overtone, respectively. The ratio of $P_{1}$ / $P_{0}$ of \target is measured as 0.776. In general, the single or double radial mode is common in HADS pulsators \citep{Breger2000} and a detailed diagram of double-mode HADS stars and metallicities are given by \cite{Petersen1996}. But the light curve variation of peak-to-peak amplitude is lower than 0.3 mag, indicating that \target is not a HADS star but a low-amplitude radial double-mode $\delta$ Scuti star, this might indicate it is necessary to revise the definition of a HADS which is, in addition, quite arbitrary.

\begin{deluxetable*}{ccccccccc}
\tabletypesize{\small}
\tablewidth{0pc}
\tablenum{3}
\tablecaption{Comparison of the characteristics of triple-mode variables \label{Tab3}}
\tablehead{
\colhead{ }&
\colhead{AC~And}&
\colhead{V823~Cas}&
\colhead{V829~Aql}&
\colhead{GSC~762-110}&
\colhead{GSC~03144-595}&
\colhead{KIC 10975348}&
\colhead{This paper}&
}
\startdata
Period F (d) & 0.7112 & 0.6690 & 0.2924 & 0.1945 &  0.2036   & 0.0977     &  0.0861\\
$F$     & 1.4059      & 1.4947 & 3.4150 & 5.1412 &  4.9099   & 10.231     &  11.614 \\
$1O$    & 1.9043      & 1.9505 & 4.5290 & 6.7284 &  6.4319   & 13.498     &  14.974 \\
$2O$    & 2.3749      & 2.4335 & 5.6640 & 8.3974 &  8.0351   &  -         &  -\\
$3O$    &  -     & - & - & - & -                             & 19.000     &  -\\
$F/1O$  & 0.7383      & 0.7663 & 0.7555 & 0.7641 &  0.7656   & 0.7579     &  0.7756  \\
$F/2O$  & 0.5920     & 0.6142  & 0.6001 & 0.6120 &  0.6112   & -         &   -   \\
$F/3O$    &  -     & - & - & - & -               &  0.5385               &   - \\
$1O/2O$ & 0.8018     & 0.8015  & 0.7997 & 0.8012 &  0.8001   & 0.7105     &  0.7885  \\
$A_F$ (mmag) & 204   & 86      & 82     & 75     &  94       & 269       &  16.5 \\
$A_{1O}/A_F$ & 0.84  & 1.51    & 1.04  & 1.00    &  0.97     & 0.0055    &  0.4864\\
$A_{2O}/A_F$ & 0.35  & 0.26    & 0.35   & 0.53   &  0.15     & -          &  -    \\
$A_{3O}/A_F$ & -  & -    & -   & -   &  -           & 0.0018              &  - \\
  \enddata
   \tablecomments{Reference \citep{1976ApJ...203..616F}, \citep{jur2006}, \citep{1998IBVS.4549....1H}, \citep{2008wils}, \citep{2016AJ....152...17M}, \citep{2021AJ....161...27Y}.}
\end{deluxetable*}

Moreover, two frequencies of radial modes have been detected in \target, enriching the sample of multiple-mode variables.
In the low amplitude $\delta$ Scuti stars, it is uncommon for all independent frequencies to be radial pulsation modes. Therefore, the study of the double radial pulsation modes of \target might shed light on the intrinsic evolution mode of $\delta$ Scuti stars. Since no low-amplitude $\delta$ Scuti stars with radial multi-pulsation modes are similar to this star, we chose six HADS for comparison. The basic pulsation parameters of \target and six triple-mode HADS from literature are lists in Table \ref{Tab3}.
The detections of those weak frequencies in \target are partly due to the high-precision observations of $Kepler$. Compared with other comparison stars, \target has the smallest amplitude. \citet{1975ApJ...200..343B} propose that the observed radial pulsation modes are related to the temperature. Fundamental pulsators mainly lie on the cool side of the instability strip, while the stars with overtones tend to be hot pulsators. \target with a overtone would have a high temperature. The two frequencies of radial modes detected in low-amplitude $\delta$ Scuti \target without non-radial modes is unusual, suggesting this star might be a post main sequence $\delta$ Scuti crossing the instability strip for the first time \citep{2010AcA....60....1P}. We suggest that detailed studies on the period ratios and seismic modeling might help to understand the nature of the double radial pulsation mode of \target. The observations of $\delta$ Scuti stars from $TESS$ also provide well-sampled photometric data, which help shed light on interiors pulsation modes of $\delta$ Scuti stars, and hence the multiple radial modes.

\subsection{Amplitude growth in \target}
\citet{2016MNRAS.460.1970B} propose that the amplitude of \target might increase somewhat with time. We then investigate the amplitude variation of the three frequencies with the highest amplitude. Following the method described by \citet{2012MNRAS.427.1418M}, we employ a non-linear least-squares fit of the frequencies to the data of each quarter using the software PERIOD04 \citep{Lenz2005}. The frequency, amplitude, and phase of the peak from the high-resolution amplitude spectrum for all data are utilized as input parameters, while the fixed frequency of each mode used for tracking amplitude and phase of the whole data set can be obtained from the least square fitting. The fixed frequency is then used to optimize the amplitude and phase of each time bin, and the same method has been applied to track the amplitude and phase for the other frequency.

\begin{figure*}
\begin{center}
  \includegraphics[width=0.87\textwidth]{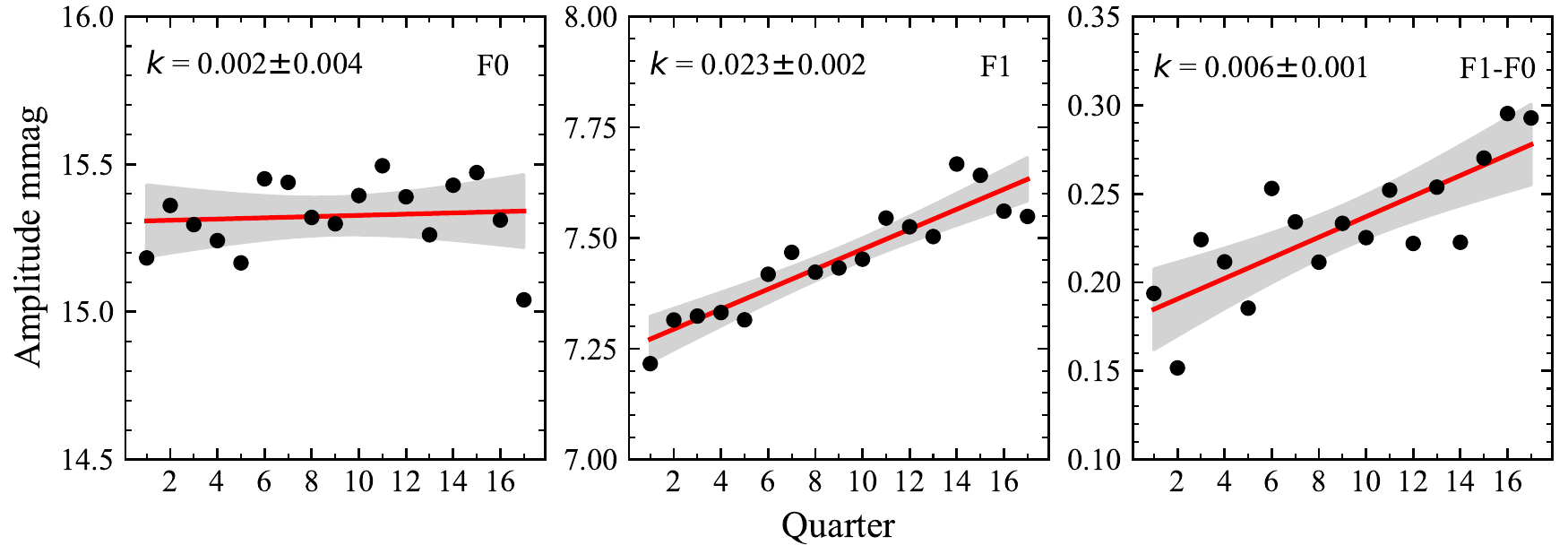}
  \caption{The amplitude variations of F1 and F1-F0 as a function of time. The red lines represent the linear fits of those frequencies using emcee (a Markov chain Monte Carlo Ensemble sampler, \citealp{2013PASP..125..306F}). Errors on amplitude are of order of a few micromagnitudes, much small than the plot symbols. }
    \label{fig:three}
\end{center}
\end{figure*}

Figure \ref{fig:three} shows slow growth in the amplitude of F1 and $f_{3}$ (F1-F0), while a relatively flat evolution in the amplitude of F0. According to \citet{2012MNRAS.427.1418M}, two potential scenarios have been proposed for the apparent amplitude growth: (1) the amplitude growth is in some way instrumental; and (2) the amplitude growth is an intrinsic or extrinsic effect to the star.

The slight variation of F1 and $f_{3}$ might be somewhat a result of systematic error from the instrument. So we check the contamination of \target from the Kepler Input Catalogue (KIC; \citep{ 2011AJ....142..112B}). The contamination reported for \target is 0.024, significantly smaller than KIC 3429637 \citep{2012MNRAS.427.1418M}, suggesting that the amplitude increasing of F1 is not an instrumental signal. Besides, \citet{2012MNRAS.427.1418M} demonstrate that it could not be an instrumental effect, as all modes would decrease or increase with the same functional form if the modulation is instrumental, and instrument trends generally affect low frequencies only.

The 4 CVn is the only one $\delta$ Scuti star with detailed studies of amplitude modulation based on ground-based telescopes. \citep{2000MNRAS.313..129B,2009AIPC.1170..410B,2016A&A...592A..97B}. \citet{2016MNRAS.460.1970B} utilize observations of $\sim$1000 $\delta$ Scuti stars from the Kepler Space Telescope \citep{2010Sci...327..977B} to investigate the amplitude modulation as well as its possible mechanism. The reason for $\delta$ Scuti stars with variable pulsation amplitudes (and/or frequencies) could roughly be classified as the intrinsic and extrinsic, i.e., physical, interior variations to the star and external effects \citep{2016MNRAS.460.1970B}. The beating from pairs (or groups) of close-frequency \citep{2002A&A...385..537B,2006MNRAS.368..571B} and non-linearity or mode coupling \citep{2016MNRAS.460.1970B} could be considered intrinsic effects.

The amplitude variation of \target may be inconsistent with beating effects since no pairs of close-frequency have been detected. \citet{2012ApJ...759...62B} propose that three frequencies within a family follow the frequency, amplitude and phase relations described in equations $\nu_{1}$ $\simeq$ $\nu_{2}$ $\pm$ $\nu_{3}$, $A_{1}$ = $\mu$$_{\rm c}$ ($A_{2}$  $A_{3}$), $\phi_1$ = $\phi_2$ $\pm$ $\phi_3$, where $A_i$ and $\phi_i$ represent the amplitude and phase of the child and parent modes, respectively, and $\mu_{\rm c}$ is the coupling factor. The $\mu_{\rm c}$ could be small values for combination frequencies from the non-linear distortion model. For \target, the $f_{3}$ is a combination frequency of F0 and F1 with a $\mu_{\rm c}$ estimated as 0.0025 which might consistent with the non-linear distortion model. However, \citet{1982AcA....32..147D} suggest that two linearly unstable low-frequency parent modes can damp a high-frequency unstable child mode once they reach the critical amplitude. We propose that the amplitude modulation of the $f_{3}$ detected in \target could not originate from the non-linearity for either of F0 nor F1 is low-frequency modes. \citet{2020MNRAS.498.1194L} utilize a different approach which is based on a Volterra expansion to describe the nonlinearities. \citet{2012MNRAS.427.1418M} propose that stellar evolution is the primary reason for the amplitude variation of modes in KIC 3429637. The amplitude variation of \target could be result from stellar evolution.

\section{SUMMARY}

We have analyzed the pulsating behavior of \target using high-precision photometric observations from the $Kepler$ mission, and eight significant frequencies are detected. While two of them are independent frequencies, i.e. F0 = 11.6141 d$^{-1}$, F1 = 14.9741 d$^{-1}$. The ratio of P1/P0 of KIC 12602250 is measured to be 0.776, suggesting that this target could be a new low amplitude radial double-mode $\delta$ Scuti star. This may indicate the modification of the definition of HADS is necessary. The low-amplitude, double-mode, and slow amplitude growth of F1 and $f_{3}$ might suggest that KIC 12602250 could be a post main sequence $\delta$ Scuti crossing the instability strip for the first time \citep{2010AcA....60....1P}. The slow amplitude growth detected in F1 might an indicator of stellar evolution. In order to confirm that the pulsation modes detected using Kepler photometry are indeed overtones and to reveal the nature of the amplitude variation found in KIC 12602250, a follow-up spectroscopic campaign and a full seismic modelling are necessary.

\acknowledgements

We thank the anonymous referee for the suggestive comments, which improved the manuscript. This research is supported by the National Natural Science Foundation of China (grant No. U2031209 and 12003020). JPG acknowledge funding support from Spanish public funds for research from project PID2019-107061GB-C63 from the 'Programas Estatales de Generaci\'on de Conocimiento y Fortalecimiento Cient\'ifico y Tecnol\'ogico del Sistema de I+D+i y de I+D+i Orientada a los Retos de la Sociedad', and from the State Agency for Research through the "Center of Excellence Severo Ochoa" award to the Instituto de Astrof\'isica de Andaluc\'ia (SEV-2017-0709), all from the Spanish Ministry of Science, Innovation and Universities (MCIU). We would like to thank the $Kepler$ science team for providing such excellent data.

\end{document}